\documentclass{article}  
\usepackage{lajolla2006}
\usepackage{graphicx}
\frompage{000} \topage{000}                                              
\def\rightmark{Systematic Study at High $p_T$}
\def\leftmark{C. Klein-B{\"o}sing}
 
\title{Systematic Study of Particle Production at High $p_T$ with\\
the PHENIX Experiment} 
\authors{
{Christian Klein-B{\"o}sing$^1$ for the PHENIX Collaboration %
}\\[2.812mm]
{\normalsize
\hspace*{-8pt}$^1$ Institut f{\"u}r Kernphysik, \\ 
48149 M{\"u}nster, Germany\\[0.2ex] 
}}
 
\abstract{ A systematic study of particle production at large
transverse momentum ($p_T$) with the PHENIX experiment at RHIC is
shown. We demonstrate that the suppression of the yield of high $p_T$
hadrons in central Au+Au collisions compared to the scaled p+p
reference persists up to highest $p_T$ for neutral pions and $\eta$
mesons. A similar suppression pattern is also observed in Cu+Cu
collisions at the same energy. In addition we present the first RHIC
results on high-$p_T$ particle production close to SPS energies.}

 
\begin{document}
 
\maketitle
\setcounter{page}{1}

\section{Introduction}\label{intro}

One of the primary goals of the PHENIX experiment at the Relativistic
Heavy Ion Collider (RHIC) at Brookhaven National Laboratory is to
study strongly interacting matter under extreme conditions. In
particular it is expected that a new state of matter, the Quark Gluon
Plasma (QGP), is formed when strongly interacting matter attains
energy densities above $\epsilon_c \approx 1$~GeV/fm$^3$. In this new
phase quarks and gluons are no longer confined into the color-singlet
nucleons, but instead can move freely over large distances.  

One possible signature that has been proposed for the creation of such
a new phase is the suppression of particles with large transverse
momenta ($p_T$) in central Au+Au collisions compared to the
expectation from scaled p+p reactions \cite{Gyu90a,Gyu94b}.  The
production of these particles is dominated by so-called hard
processes, parton-parton interactions with large momentum transfer
$Q^2$ ($\propto p_T^2$), and the subsequent fragmentation of partons
into observable particles, which are observed in elementary collisions
as \emph{jets} of particles produced along the direction of the
scattered partons. In heavy-ion collision these hard processes occur
in the early stage of the reaction, before the hot and dense medium is
formed, thus providing an ideal probe for the later stage.

The first step when studying medium influences on particle production
is to compare the production at high $p_T$ in heavy ion collision to
the expectation for the QCD vacuum: The reference provided by the
measurement in p+p collisions, scaled with the nuclear overlap
function ($\langle T_{AB} \rangle$) for a given centrality in A+B
reactions. This defines the nuclear modification factor:
\begin{eqnarray}
\label{eq:RAA}
R_{AB} = \frac{d^2N_{AB}/dy dp_{T}}
{\langle T_{AB} \rangle \cdot d^2 \sigma_{pp}/dy dp_{T}},
\end{eqnarray}
where $\langle T_{AB} \rangle$ is related to the number of binary
nucleon-nucleon collisions $N_{coll}$ via $\langle T_{AB} \rangle
\approx \langle N_{coll} \rangle /\sigma_{inel}$. They can be computed
together with the number of participating nucleons $N_{part}$ in a
Glauber calculation.

\begin{figure}[!thb]
\insertplot{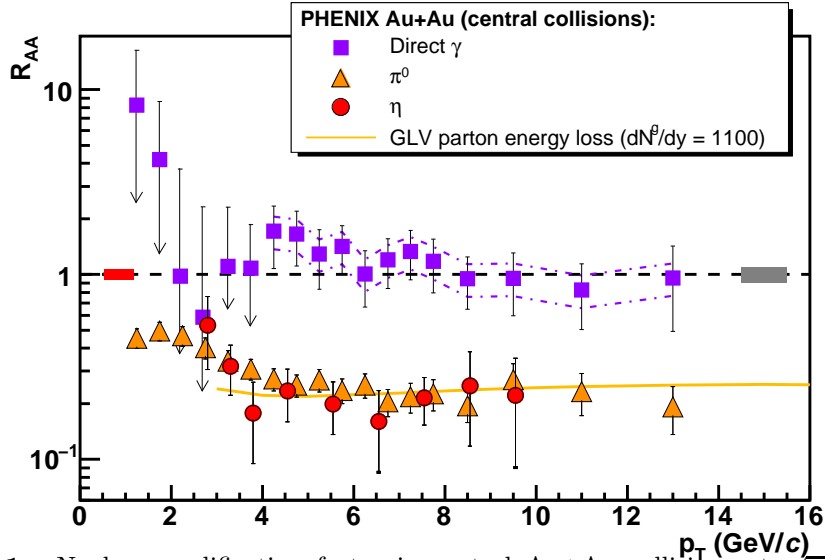}
\vspace*{-1cm}
\caption[]{Nuclear modification factor in central Au+Au collisions at
$\sqrt{s_{NN}} = 200\mbox{~GeV}$ for direct photons, $\pi^0$s and
$\eta$s \cite{Adler:2006hu}. The hadrons are suppressed as expected by
the energy loss calculation in a dense partonic medium
\cite{Vit02a,Vitev:2004bh}.}
\label{fig1}
\end{figure}

At low $p_T$ ($< 2\mbox{~GeV}/c$) the particle production is dominated
by soft processes which scales approximately with $N_{part}$ but at
sufficiently large $p_T$ hard scattering dominates and the nuclear
modification factor should be unity, any deviation from unity
indicating an influence of the medium. Indeed one of the major
discoveries at RHIC is the suppression of particle production by a
factor of five in central Au+Au collisions (see Fig.~\ref{fig1} and
\cite{Adc02a,Adl03d,Adl04a,Ada03b}), while no suppression is observed
in peripheral reactions. This suppression can be explained by the
energy loss of hard scattered quarks via induced gluon bremsstrahlung
in a medium with high parton density (see e.g. \cite{Gyu03a}), which
is supported by two key observations.

\begin{itemize}
\item In d+Au reactions only a slight Cronin enhancement is observed
but no suppression \cite{Adl03f,Kle04a}: Initial state effects,
such as the Color Glass Condensate, cannot be responsible for the
suppression in central Au+Au collisions.
\item The nuclear modification factor for direct photons is consistent
with unity for all centralities in Au+Au (see Fig.~\ref{fig1} and
\cite{Adl05a}): The direct control in Au+Au reactions that the hard
scatterings occur at the expected rate, since photons do not interact
strongly with the medium.
\end{itemize}

However, to distinguish between various models for the parton-energy
loss and to map out the properties of the medium, more differential
studies of high-$p_T$ particle production are needed. This can involve
the study of heavy quarks, jet correlations, parton energy-loss with
respect to the event reaction plane, dependence on particle species,
system-size and energy dependence. Here we will concentrate mainly on
the system-size and energy dependence of the nuclear modification
factor and will limit ourselves to the measurement of identified
particles above $p_T \approx 6~\mbox{GeV}/c$, a region where the
difference between charged hadrons and neutral pions, which motivated
the picture of quark-recombination from a thermal source as dominant
particle source is no longer important \cite{Adl04a}.

\section{Data Analysis}

The data presented here were obtained during the first five years of
operation of the PHENIX experiment at RHIC. The primary detector
employed in the measurement of neutral pions, $\eta$ mesons and direct
photons was the Electromagnetic Calorimeters (EMCal) \cite{Aph03a}. In
addition, the PHENIX zero-degree calorimeters and the two beam-beam
counters (BBC's) were used for triggering, vertex and centrality
determination. The EMCal is located at a radial distance of
$\sim$5.1~m and covers $\left|\eta\right| < 0.35$ in pseudo-rapidity
and $\pi$ radians in azimuth.  It is divided into eight sectors: six
sectors lead-scintillator sandwich calorimeter (PbSc) and two sector
lead-glass Cherenkov calorimeter (PbGl), which provides an excellent
internal cross-check for the final results.

\begin{figure}[!thb]
\insertplot{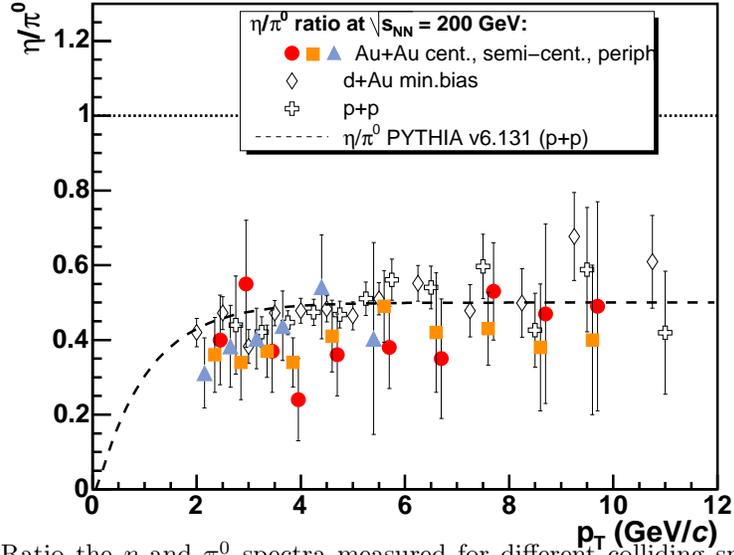}
\vspace*{-1cm}
\caption[]{Ratio the $\eta$ and $\pi^0$ spectra measured for different
colliding species and centralities at $\sqrt{s_{NN}} = 200\mbox{~GeV}$
\cite{Adler:2006hu}. All measurements are consistent with each other
and with a PYTHIA calculation \cite{Sjo00a}.}
\label{fig2}
\end{figure}

Neutral pions and $\eta$s are reconstructed with an invariant mass
analysis of photon pairs in the EMCal. The combinatorial background is
determined via an mixed event technique, where photons from different
events with similar topology (centrality, collision vertex etc.) are
combined to obtain the background from uncorrelated photon pairs.
Direct photons are measured by reconstructing the inclusive photon
yield and comparing to the photons expected from hadronic decays, any
excess above the decay background is a signal of direct photons (see
also \cite{Adl05a,Zau06a}).

\begin{figure}[!thb]
\insertplot{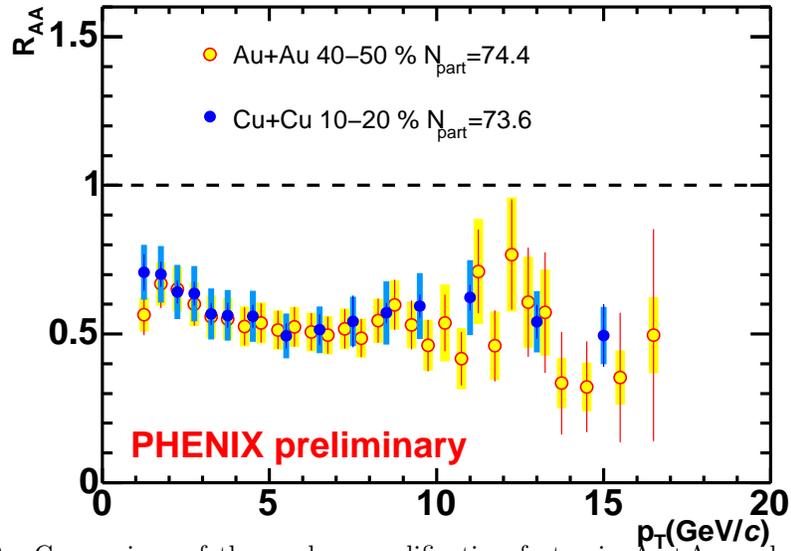}
\vspace*{-1cm}
\caption[]{Comparison of the nuclear modification factor in Au+Au and
Cu+Cu collisions at $\sqrt{s_{NN}} = 200\mbox{~GeV}$ for $N_{part}
\approx 74$.}
\label{fig3}
\end{figure}

\begin{figure}[!thb]
\insertplot{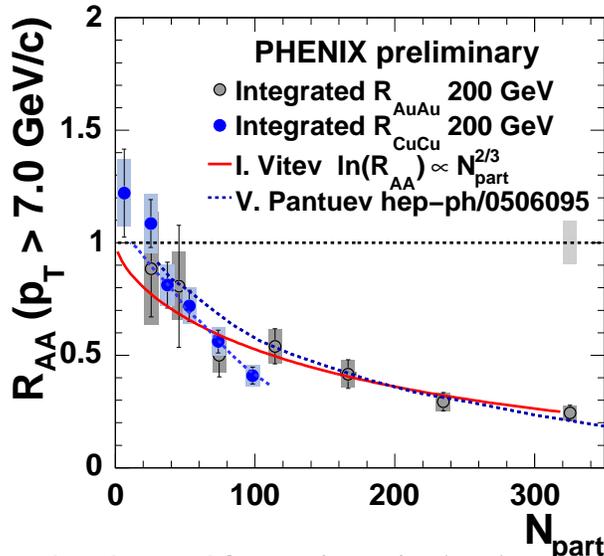}
\vspace*{-1cm}
\caption[]{Integrated nuclear modification factor for Au+Au and Cu+Cu
collisions at $\sqrt{s_{NN}} = 200\mbox{~GeV}$ compared to a scaling
with $\ln R_{AA} \propto N_{part}^{2/3}$ \cite{Vitev:2005he} and to a
calculation with surface/volume effects \cite{Pantuev:2005jt}.}
\label{fig4}
\end{figure}

\section{High $p_T$ $\eta$ production}

The fact that neutral pions and charged hadrons show the same amount
of suppression above $p_T \approx 5\mbox{~GeV}/c$ indicates that the
jet-quenching mechanism seems not to depend on the identity of the
(light-quark) hadron \cite{Adl04a}. This is expected when the
jet-quenching depends only on the energy loss of the parent light
quark and not on the nature of the final leading hadron which is
produced in the same universal fragmentation process as in hadron
production in the vacuum, i.e. in p+p collisions.

However, a similar nuclear modification factor for neutral pions and
charged hadrons does not provide a very strong argument for energy
loss on the partonic level, since the yield of unidentified hadrons
above $p_T \approx 5~\mbox{GeV}/c$ is dominated by charged pions
\cite{Adl04a}. PHENIX has measured the production of $\eta$ mesons in
Au+Au, d+Au and p+p collisions at $\sqrt{s_{NN}} = 200\mbox{~GeV}$
which allows to study the effects of hot and cold nuclear matter with
an additional probe of identified high-$p_T$ particles and to compare
with the results of neutral pion and direct photon production \cite{Adler:2006hu}. 

This is done in Fig.~\ref{fig1} for the most central collision in
Au+Au. It is clearly seen that the nuclear modification factor for
$\eta$ mesons shows the same pattern as for $\pi^0$s, a suppression of
a factor of five and the same constancy at high $p_T$. These features
are well described by the energy-loss calculation shown in
Fig.~\ref{fig1} for a medium with density $dN^g/dy = 1100$
\cite{Vit02a}. As already mentioned above, direct photons show no
suppression at high $p_T$ which nicely illustrates the insensitivity
of photons to the produced medium of quarks and gluons and provides
the \emph{in situ} confirmation that hard scatterings occur at the
expected rate in central Au+Au reactions.

A different approach to compare the production of $\pi^0$s and $\eta$s
at high $p_T$ is to study the dependence of the ratio $\eta/\pi^0$ on
collision species and centrality. Since $\eta$s and $\pi^0$s are
reconstructed within the same data set via their decay into two
photons, many systematic uncertainties cancel. The result is shown in
Fig.~\ref{fig2} for d+Au, p+p reactions and for three different
centralities in Au+Au together with a PYTHIA calculation. The ratios
for all colliding species and centralities are consistent within the
errors, again demonstrating that the energy loss occurs at the
partonic level and the fragmentation process is not strongly affected
by the medium.

\section{System Size Dependence}

The study of different system sizes in general is motivated by the
fact that a smaller colliding system (e.g. Cu with $A = 63$) allows for
a more precise discrimination for smaller values of $N_{part}$,
corresponding to peripheral collisions with Au ($A = 197$). In
addition it allows for the study of the effects of collision geometry,
since at the same $N_{part}$ the overlap region for the smaller
Cu-system is more spherical and the surface/volume ratio is
smaller. For these reasons Cu+Cu collisions were studied in the fifth
year of physics running (2004/2005) at RHIC.

\begin{figure}[!thb]
\insertplot{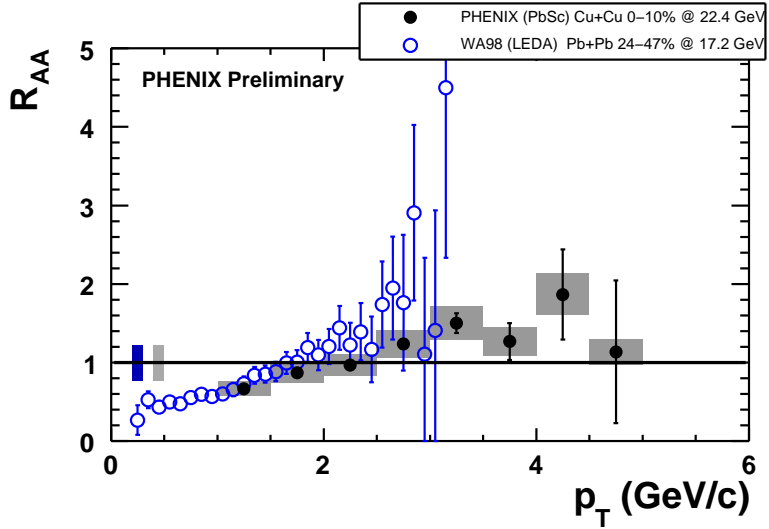}
\vspace*{-1cm}
\caption[]{Nuclear modification factor for mid-central Pb+Pb
collisions ($N_{part} = 132$, $\sqrt{s_{NN}} = 17.3$~GeV) measured by
WA98 at CERN-SPS and for central Cu+Cu collisions ($N_{part} = 140$,
$\sqrt{s_{NN}} = 22.4$~GeV) measured by PHENIX at RHIC.}
\label{fig5}
\end{figure}

At the same number of participating nucleons and at the same
$\sqrt{s_{NN}}$, corresponding to a similar energy density, the
nuclear modification factor for neutral pions is basically identical
for Au+Au and Cu+Cu collisions as shown in Fig.~\ref{fig3}. This is
also illustrated in Fig.~\ref{fig4}, where the integrated nuclear
modification is shown for different centralities. It can be seen that
$R_{AA}$ is very similar for Au+Au and Cu+Cu and approximately follows
the scaling with $\ln R_{AA} \propto N_{part}^{2/3}$ proposed in
\cite{Vitev:2005he}. However, it seems like the $N_{part}$ dependence
reveals a slightly different slope for Cu+Cu and Au+Au, pointing to
geometrical effect on the nuclear modification factor. In
\cite{Pantuev:2005jt} a model was developed which assumed that only
jets originating from a certain depth below the surface of the
collision region can be observed, the predicted $N_{part}$ dependence
(see Fig.~\ref{fig4}) shows a remarkable agreement with the
measurement. Though the interpretation of this observation is not yet
clear, it becomes obvious that geometry effects, such as a surface
bias, need to be understood when comparing the medium properties for
different colliding species.

\section{Energy Dependence}

The suppressed production of hadrons at high-$p_T$, observed by the
RHIC experiments at $\sqrt{s_{NN}} = 200~\mbox{GeV}$ and 130~GeV, also
triggered the discussion, if and to what extent jet-quenching plays
already a role at SPS energies and how the energy dependence
evolves. After a re-evaluation of the existing SPS data with an
improved p+p reference \cite{d'Enterria:2004ig}, it appears that some
amount of jet-quenching may already be present in central heavy-ion
collisions at $\sqrt{s_{NN}} \approx 20~\mbox{GeV}$. To finally answer
this question, the first step has been taken by the PHENIX experiment
by measuring the production of neutral pions in Cu+Cu collisions at
$\sqrt{s_{NN}} \approx 22.4~\mbox{GeV}$ up to $p_T \approx
5~\mbox{GeV}/c$. Though the measurement of the p+p reference within
the same experiment is still missing at the moment, there exists a
wealth of data for p+p $\rightarrow \pi^0/\pi^\pm + X$ in the range
$\sqrt{s_{NN}} = 21.7-23$~GeV which allows to construct a p+p
reference with a systematic uncertainty of $\approx$ 20\%.

The resulting nuclear modification shows no strong variations with
centrality, instead it indicates Cronin-enhancement. The comparison of
the PHENIX result for Cu+Cu and the WA98 result in Pb+Pb for similar
$N_{part}$ is shown in Figure~\ref{fig5}. Both measurements agree well
where they overlap and the PHENIX measurement nicely extends to higher
$p_T$. What is still missing is the measurement of the p+p reference
within the same experiment, to improve the significance of the result
in Cu+Cu and also a measurement of Au+Au collisions at the same
colliding energy to reach higher energy densities than in Cu+Cu
reactions.

\section{Conclusion}

We have shown a systematic study of high $p_T$ particle production by
comparing the nuclear modification factor $R_{AA}$ for different particle
species, colliding species and center-of-mass energies.  We
demonstrated that the $\pi^0$ and $\eta$ yield in central Au+Au at
$\sqrt{s_{NN}} = 200$ collisions show a similar suppression pattern,
indicating that the suppression is happening at the partonic
level. $R_{AA}$ for different Au+Au and Cu+Cu reactions shows a very
similar dependence on $p_T$ and $N_{part}$, with some indications that
geometric effects can become important when comparing the most central
Cu+Cu to Au+Au collisions at similar $N_{part}$. We have also shown
the first measurement of high-$p_T$ particle production near SPS
energies ($\sqrt{s_{NN}} = 22.4$~GeV) at RHIC which agree with previous
measurements by WA98 and extended the measurement to higher
$p_T$. The reference measurement at RHIC with p+p collisions at the
same energy is still needed.

\bibliographystyle{lajolla2006} 
\bibliography{Klein-Boesing_2006-LaJolla}

\vfill\eject
\end{document}